\documentclass[conference]{IEEEtran}

\usepackage{amsmath}
\usepackage{amssymb}
\usepackage[linesnumbered,ruled,vlined]{algorithm2e}
\usepackage{bm}
\usepackage{multirow}
\usepackage[utf8]{inputenc}
\usepackage{graphicx}
\usepackage{booktabs}
\usepackage[draft]{hyperref}
\DeclareGraphicsExtensions{.pdf,.jpeg,.png}

\begin{document}

\title{When Rotation Bounds Nothing: Streaming Image Classifiers to Field Nodes Under Physical Capture}

\author{\IEEEauthorblockN{1\textsuperscript{st} Daniel Gilkarov}
\IEEEauthorblockA{\textit{Dept.\ of Computer and Software Engineering} \\ \textit{Ariel Cyber Innovation Center}\\ Ariel University, Israel\\
daniel.gilkarov1@msmail.ariel.ac.il}
\and
\IEEEauthorblockN{2\textsuperscript{nd} Ran Dubin}
\IEEEauthorblockA{\textit{Dept.\ of Computer and Software Engineering} \\ \textit{Ariel Cyber Innovation Center}\\ Ariel University, Israel\\
rand@ariel.ac.il}
}

\maketitle

\begin{abstract}
Field nodes run image classifiers locally, exposing a first-class asset, the trained model, to
theft. Trusted-execution enclaves, collaborative-inference splits, and encrypted storage assume
the device owner is the adversary from deployment. Here the operator owns the node, and an
adversary seizing it reads only what cold-boot capture finds resident. One answer stores nothing:
a chunkable multi-exit model streams into volatile memory, rotating to bound residency. We
characterise the capture envelope on real LTE traces across two victims of one architecture,
against passive and adaptive adversaries. Neither survives a capable adversary; recovery cost
alone differs. Against a from-scratch victim the uncaptured stages must be learned: the adaptive adversary gains
most at shallow seizures, ends the CIFAR-10 envelope at every rotation, and stays rotation-bounded
on CIFAR-100, from $42.6\%$ of usable seizures at $5$\,s to $86.8\%$ at $30$\,s. Against the same architecture sliced from a public pretrained backbone, the
adversary downloads those stages, recovering $\geq\!90\%$ of maximum accuracy at all depths,
taking usable CIFAR-100 seizures from $0.0\%$ to $97.3\%$ at $5$\,s, given the victim's full
labelled training set. Delivery-side secrecy raises a seizure's price; a public copy of the
withheld stages erases it.
\end{abstract}

\section{Introduction}
\label{sec:intro}
Consider a fleet of consumer or in-vehicle units that run an image classifier but may not keep it on
local storage, because a licence term forbids persisting it to disk or because it rotates faster
than a cached copy stays fresh. Each unit re-fetches the model into volatile memory every session,
over a cellular link whose throughput swings with handovers and congestion. Delivery, not
execution, binds there, and a monolithic artifact that yields nothing until its last byte
arrives cannot be served at all when the link stalls. A chunkable multi-exit
model~\cite{rahmath2024, msdnet2018} instead runs from its first chunk and improves as bandwidth
permits, so the device is operational from a small resident prefix.

That resident prefix is also a security exposure: the model lives only in RAM and the operator
owns the node, so the adversary who matters is not the device owner but one who physically
seizes it and reads its memory by cold-boot or DMA capture~\cite{lest_we_remember},
inverting the digital-rights-management (DRM) assumption of an owner adversarial from
deployment (\S\ref{sec:threat}).

Rotating the model and re-streaming it from the first chunk turns recovery into a race against a
bounded residency window, and we characterise the resulting envelope on real LTE traces across two
victim models of one architecture. Which victim it is decides the answer, because the threat model
concedes the architecture to the adversary and a public backbone concedes the withheld weights with
it. That conflict is the paper's finding: the measurement prices recovery in each regime, and names
the deployment class in which it is free.

\subsection{Contributions}
\begin{enumerate}
  \item \textbf{A field-capture threat model, distinct from DRM.} Prior on-device model protection
        assumes the device owner is the adversary and holds the device permanently. We assume the
        defender owns the device and the adversary's access begins only at seizure, inverting the
        DRM setting's core assumptions of ownership, access, and time budget, and making recovery a
        race against a bounded residency window (\S\ref{sec:threat}).
  \item \textbf{Two victim regimes over one byte axis.} We build the same $5$-exit frontier twice,
        joint-trained from scratch and sliced from a public ImageNet-1k ResNet-18 and fine-tuned. The
        payload axis does not move between them, so only accuracy does, and the pretrained
        construction is the better model ($a_{\max}$ $0.9256\!\rightarrow\!0.9428$ on CIFAR-10,
        $0.6951\!\rightarrow\!0.7726$ on CIFAR-100). Sharing that byte axis, the two passive
        envelopes coincide at every swept bar but $\tau{=}0.8$, where the pretrained victim's higher
        bar leaves fewer seizures usable on both datasets (\S\ref{sec:results-capture}); and that
        passive adversary is only a naive baseline: neither regime survives a capable one that learns
        those stages against the from-scratch victim or downloads them against the pretrained
        (item~3), and the passive bound is not free: \S\ref{sec:results-resident} prices it in
        accuracy and energy.
  \item \textbf{The public prior, measured.} An adversary that initialises the withheld stages from
        the same public weights the defender started from, then fine-tunes exactly as before,
        recovers at least $90\%$ of $a_{\max}$ at every seizure depth on both datasets and takes the
        CIFAR-100 envelope from $0.0\%$ to $97.3\%$ usable seizures at a $5$\,s rotation. An identical
        adversary that must learn those stages, again on the same pretrained victim, clears the bar
        in $2$ of $135$ eligible cells against $50$, so the cause is the public prior and not an
        easier victim (\S\ref{sec:results-thief}). On
        CIFAR-10 the collapse needs $5000$ labelled samples; on CIFAR-100 it needs the victim's
        entire training set. The consequence is that this deployment pattern and this class of
        defence are incompatible.
\end{enumerate}

\section{Threat Model: Field Capture versus DRM}
\label{sec:threat}
The setting is a node the operator owns and controls: a fielded sensor or in-vehicle unit that
streams an image classifier into volatile memory each session, writes no weights to disk, and
reports telemetry over a cellular link. The asset is the model. The adversary physically seizes a
node; the seizure recovers the prefix resident when the node leaves the operator's service, on
either side of detection. A node seized live stops reporting, and revocation bounds
its further streaming; one seized after a takedown cannot complete the next rotation's re-stream
over the severed link, so by design (\S\ref{sec:rotation}) its residency only shrinks. The
envelope thus upper-bounds both orderings, and \S\ref{sec:eval} samples the seizure instant
across live operation, independently of the rotation phase, so it reports the phase mean;
\S\ref{sec:results-capture} measures the per-rotation maximum an adversary timing seizures to the
periodic downlink burst would read instead. Cold-boot and DMA techniques make whatever model bytes remain in RAM
extractable once the device is in hand~\cite{lest_we_remember}. We consider two adversaries: a
\emph{passive} one who uses the seized bytes as they are, and an \emph{adaptive} one who
reconstructs the withheld stages from a captured prefix and fine-tunes them, granted
in-distribution labelled data up to the victim's full training set and its exact training recipe,
more than a field-capture adversary plausibly holds (\S\ref{sec:results-thief}). The architecture is conceded to both, since the backbone is a published
design and treating its topology as secret would be security through obscurity. Whether the withheld
\emph{weights} are equally obtainable is the question \S\ref{sec:results-thief} settles.

This threat differs in every dimension from the digital-rights-management setting, as
Table~\ref{tab:threat} sets out. Because the adversary's access is bounded in time in either
ordering, recovery becomes a race rather than a certainty---which is why an envelope, and not a
defence, is the object we measure.

\begin{table}[tb]
\centering
\caption{The field-capture threat model inverts the DRM assumptions that on-device model protection
has almost always made.}
\label{tab:threat}
\setlength{\tabcolsep}{4pt}
\footnotesize
\begin{tabular}{p{0.30\columnwidth}p{0.28\columnwidth}p{0.30\columnwidth}}
\toprule
 & DRM (prior art) & Field capture (ours) \\
\midrule
Device owner & adversary & defender \\
Adversary access & permanent, from deployment & only after seizure \\
Time budget & unlimited & bounded by detection, revocation \\
Capture detectable & no capture event & yes: node stops reporting \\
May destroy device state & no (customer's device) & in principle; channel severed \\
Link live during attack & yes & not required; read offline \\
\bottomrule
\end{tabular}
\end{table}

\section{Related Work}
\label{sec:related}
On-device model protection has almost always taken the device owner to be the
adversary. Trusted-execution approaches run inference inside an enclave---Intel
SGX~\cite{costan2016intel}, ARM TrustZone~\cite{pinto2019demystifying}---so that a
malicious owner cannot read the weights. ShadowNet~\cite{shadownet2023} partitions a CNN to keep
its linear layers inside the TEE, and TEESlice~\cite{teeslice2024} protects a model's sensitive
slices even against an owner holding a pre-trained copy. Collaborative-inference splits
(Neurosurgeon~\cite{Kang2017}, SPINN~\cite{laskaridis2020}) keep early layers on the device and the
remainder in the cloud, again to withhold it from an untrusted edge. Each assumes
permanent adversarial access from deployment and a device the defender may not touch---the
DRM setting of \S\ref{sec:threat}. We invert it: the defender owns the node, and the adversary
appears only after a physical seizure. The two classes share a failure condition
rather than differing in one: the adversary \S\ref{sec:results-thief} models, which already holds the
pre-trained weights the defender built on, is the threat TEESlice names for
itself. Encrypted on-device storage persists a key-protected artifact and rests on where its key is
held; this design persists nothing and rests on the residency window, so a head-to-head would
report those assumptions rather than the mechanisms.
NNSplitter~\cite{nnsplitter2023} withholds a chosen parameter set with no public counterpart,
rather than a slice of a downloadable backbone, so the condition \S\ref{sec:results-thief}
identifies does not arise there. The seizure adversary itself is classical: sensor-network
security has treated defender-owned nodes as subject to capture since Eschenauer and
Gligor~\cite{eschenauer2002key}, and FIPS 140-class tamper response meets it with triggered
zeroisation. New are the asset and the bound: residual value that is a graded
accuracy-versus-bytes curve, not a binary key compromise, and probabilistic residency under
rotation, not triggered erasure; the on-device extraction taxonomy~\cite{nayan2024sok} assumes a
persistent model file, permanently held.

The mechanism we characterise is model streaming. Multi-exit and early-exit
networks~\cite{msdnet2018, rahmath2024} make a model runnable from a prefix, partial and
simultaneous model download (NAS~\cite{nas_osdi2018}, SLIDE~\cite{slide2025}) schedule model bytes
over a link, and quantization~\cite{jacob2018quantization} shrinks a model while leaving it
monolithic. This work treats chunkability and streaming as ways to cut delivery cost or latency,
not as a security property. We reuse the multi-exit primitive unchanged and ask what its volatile,
rotated delivery surrenders to a field adversary, a question this literature does not pose.

\section{System Design}
\label{sec:system}

\subsection{Chunkable representation}
\label{sec:repr}
The delivered model is a chunkable network with an ordered set of \emph{chunks}
$c_1,\dots,c_n$ of cumulative on-wire size $b_1<\dots<b_n$. After receiving the prefix
$c_1\dots c_k$ the device runs a valid forward pass with served accuracy $a_k$, monotone in $k$.
We instantiate this with a co-trained multi-exit backbone~\cite{rahmath2024, msdnet2018}, in which
exit $k$ attaches a lightweight head to an intermediate stage, so a partial download yields a
usable classifier and not a broken fragment. The representation is prior art
(\S\ref{sec:related}); the system requires only the chunk interface $(b_k, a_k)$.

\subsection{Streaming into volatile memory}
The server holds the published model; the client verifies each received chunk, deserializes it into
RAM without writing to disk, and serves at the new depth as soon as the chunk arrives. A
telemetry-estimated link capacity $C(t)$ and a per-session deadline $T$ give a byte
budget $B(t)=\alpha\,C(t)\,T$ with a jitter discount $\alpha\!\le\!1$; a greedy controller streams
the deepest deliverable prefix $k^\star=\max\{k: b_k\le B(t)\}$, smoothing the capacity estimate so
transient dips do not thrash the served depth. This
greedy streaming policy is the mechanism whose recovery we measure, and
\S\ref{sec:results-capture} shows it is a tight upper bound on a prefix-constrained optimal
scheduler.

\subsection{Rotation and the residency window}
\label{sec:rotation}
The operator rotates the served model with period $R$. On each rotation the device discards the
resident model (whether or not the re-stream that follows can complete) and re-streams from
$c_1$, so a node holds at most one prefix at a time and its
depth is set by how much the link delivered since the last rotation. A seizure at an arbitrary
instant therefore recovers only the prefix resident in the current rotation, which is why the
residency window $R$, and not the model's total size, bounds what a seizure obtains. Rotation is a
\emph{passive} consequence of never persisting the model and refreshing it periodically; it is not
a tamper-response or self-destruct mechanism, and no active erasure is triggered by the seizure
itself. An operator able to signal the node could zeroise it, but a takedown severs exactly the
channel such a command needs; the rotation discard is the same erasure made
detection-independent. It bounds recovery only probabilistically (a seized node is unlikely to
hold a deep prefix when rotation is fast), not by destroying state on detection.

\section{Evaluation}
\label{sec:eval}

\subsection{Setup}
\label{sec:setup}
\textbf{Models.} The chunkable representation is a $5$-exit multi-exit CNN with a ResNet-style
backbone; exit $k$ has a measured on-wire size $b_k$ and served accuracy $a_k$
(Table~\ref{tab:frontier}). We build two victim models from it. The \emph{from-scratch} victim is
joint-trained end to end; the \emph{pretrained} victim copies a torchvision ImageNet-1k ResNet-18
into stages $1$--$4$ and fine-tunes them, and a third arm freezes those stages and trains the exit
heads alone. Unit $0$ uses a $3\!\times\!3$ stride-$1$ CIFAR stem trained from scratch in all arms, since the
ResNet-18 $7\!\times\!7$ stride-$2$ stem would destroy a $32\!\times\!32$ input. All three share the
architecture, so the byte axis $b_1\dots b_5$ does not move (differing by at most $13$ bytes at
$c_1$ and not at all after) and only accuracy does, so every analytic below re-runs on either
frontier with no delivery number changing.
A second, far smaller $5$-exit backbone (MobileNet-style, from scratch) on CIFAR-10 is sized in
\S\ref{sec:results-capture}. The monolithic baseline is the same
backbone delivered whole, whose smallest fully usable form is a near-lossless int8 quantization at
$11.31$\,MB decimal against the $44.79$\,MB fp32 payload.
\textbf{Datasets.} CIFAR-10 and CIFAR-100; ImageNet-1k enters only as the pretrained backbone's
source task.
\textbf{Network.} Real captured LTE driving traces replayed with Mahimahi~\cite{mahimahi}
(provenance: Sprout~\cite{sprout2013}): AT\&T ($120.0$\,s, $68.4$\,MB, mean $4.56$\,Mbps) and
T-Mobile ($474.6$\,s, $757.6$\,MB, mean $12.77$\,Mbps).
\textbf{Protocol.} Each frontier in Table~\ref{tab:frontier} is $3$ independent training seeds
(mean). The capture, residency, and exchange results are analytic functions of one measured frontier
and the replayed traces under rotation $R$: we score the device's time-averaged served accuracy
$\bar a/a_{\max}$ together with, over seizures placed across
the trace, the recovered accuracy and the fraction of seizures clearing the usability bar
$\tau{=}0.8$ of $a_{\max}$. The bar is set by the $a_{\max}$ of the regime under test, so it rises
with it, from $0.7405$ to $0.7542$ on CIFAR-10 and from $0.5561$ to $0.6181$ on CIFAR-100. Delivered
bytes come from integrating the measured trace capacity, which assumes the link hands over bytes at
trace rate. A real transport delivers strictly less, so every envelope number is an upper bound on
what a seizure recovers, as \S\ref{sec:results-mmlink} confirms.
\textbf{Training.} One configuration serves every exit, both datasets and both victim regimes: $30$
epochs of SGD, learning rate $0.1$, momentum $0.9$, batch $128$, weight decay $5\!\times\!10^{-4}$,
width multiplier $1.0$, cosine schedule after a linear warmup. Nothing is tuned per exit, so an
accuracy difference between exits is attributable to the exit and not to its hyperparameters.
Per-exit standard deviations over $3$ seeds are at most $0.0057$ in both victim regimes.

\subsection{The frontier in three regimes}
\label{sec:results-frontier}
Practitioners build on published backbones, so whether a chunkable model must be trained from
scratch decides whether the method is deployable; it need not be. Table~\ref{tab:frontier}
reports the same $5$-exit frontier in three regimes over one byte axis. The pretrained victim
is the stronger model: $a_{\max}$ rises from $0.9256$ to $0.9428$ on CIFAR-10 and from $0.6951$ to
$0.7726$ on CIFAR-100 over the same payload, at no peak-accuracy cost. It does cost accuracy at the shallow exits, where $c_2$ falls
from $0.7799$ to $0.7414$ on CIFAR-10 and from $0.4001$ to $0.3216$ on CIFAR-100, far outside seed
variance. That steeper frontier, together with a
usability bar that rises with $a_{\max}$, is what tightens the passive envelope in
\S\ref{sec:results-capture}. A linear probe on the frozen backbone (frozen-probe rows,
Table~\ref{tab:frontier}) does not support this frontier, so fine-tuning answers the concern, not
the strictest never-touch reading.

A prefix is cheap to run: $0.617$\,MB predicts in $1.08$\,s on AT\&T, peak memory $34.9$ to $88.4$\,MB at batch $1$.

\begin{table}[tb]
\centering
\caption{The chunkable frontier in three regimes (top-1, 3-seed mean). The architecture is the same
in all three, so the payload axis does not move and only accuracy does. The from-scratch and
pretrained victims are the two the envelope results use; the frozen probe is a negative result.}
\label{tab:frontier}
\setlength{\tabcolsep}{3.5pt}
\footnotesize
\begin{tabular}{llccccc}
\toprule
dataset & regime & $c_1$ & $c_2$ & $c_3$ & $c_4$ & $c_5$ \\
\midrule
\multirow{4}{*}{CIFAR-10}
 & payload (MB)  & 0.013 & 0.617 & 2.740 & 11.169 & 44.794 \\
 & from scratch  & 0.3342 & 0.7799 & 0.9061 & 0.9251 & 0.9256 \\
 & pretrained    & 0.3286 & 0.7414 & 0.9028 & 0.9397 & 0.9428 \\
 & frozen probe  & 0.3305 & 0.4114 & 0.4864 & 0.5541 & 0.6143 \\
\midrule
\multirow{4}{*}{CIFAR-100}
 & payload (MB)  & 0.037 & 0.664 & 2.833 & 11.355 & 45.165 \\
 & from scratch  & 0.1052 & 0.4001 & 0.5839 & 0.6857 & 0.6951 \\
 & pretrained    & 0.0960 & 0.3216 & 0.6033 & 0.7474 & 0.7726 \\
 & frozen probe  & 0.0967 & 0.1265 & 0.1644 & 0.2276 & 0.4466 \\
\bottomrule
\end{tabular}
\end{table}

\subsection{Depth-adaptive scheduling under real LTE}
\label{sec:results-traj}
We validate the depth-adaptive controller on the replayed real LTE
traces; the result is null. A closed-loop controller, an EWMA-only variant, a naive deepest-now
policy, and a static fixed-depth ($c_2$) baseline tie on served accuracy near $0.86$ at $R{=}30$\,s
(AT\&T, from-scratch), while robustMPC~\cite{robustmpc2015} trails at $0.844$. The static baseline
dominates on stability ($0.1$ depth changes per $100$ ticks against the closed loop's $3.4$ and
naive tracking's $13.6$), so adaptation helps only against naive tracking, not against static.

\subsection{The passive capture envelope}
\label{sec:results-capture}
A passive adversary runs the seized bytes unchanged, so what it recovers and what the device
serves are one quantity read from one resident prefix, and faster rotation lowers both together.
The two passive series of Fig.~\ref{fig:collapse} give the usable-seizure fraction on the AT\&T
trace, one per victim regime. The pretrained victim is the deployable one and the operating regime
for the rest of this section; at $R{=}5$\,s it leaves $11.8\%$ of CIFAR-10 seizures usable and no
CIFAR-100 seizure usable at all, against $69.2\%$ and $11.2\%$ for the from-scratch victim.

The envelope is set by the product of link rate and rotation period, not rotation alone: on AT\&T
no seizure recovers the whole $44.79$\,MB artifact up to $R{=}60$\,s ($P(\text{full}){=}0.000$),
which even $60$\,s of a $4.56$\,Mbps link cannot deliver, whereas on the $2.8\times$ faster T-Mobile
link complete recovery is non-zero from $R{=}10$\,s. Model size settles the question outright: the
MobileNet frontier ($2.287$\,MB in total, from scratch, with a usable exit at $0.19$\,MB) is not
protectable at any rotation we test, with $P(\text{full}){=}0.826$ at $R{=}30$\,s even on the slower
link, because that model arrives whole inside a single residency window.

The design's security cost shows against its own monolithic baseline. On the from-scratch victim the
near-lossless int8 quantization of \S\ref{sec:setup} (within $0.32$\,pp of fp32 on both datasets,
$11.31$\,MB) yields nothing until its last byte, so $P(\text{usable}){=}P(\text{full})$: no CIFAR-10
seizure is usable at $R{=}5$\,s ($0.000$ against $0.692$ for the chunkable frontier), and $0.387$ at
$R{=}30$\,s against $0.942$. At $\approx\!11.3$\,MB the monolith delivers $a_{\max}$ where the
chunkable $c_4$ ($11.169$\,MB) delivers $0.9251$: the same bytes for the same accuracy, opposite
capture profiles. Chunkability buys availability from the first chunk and pays in exposure.

An adversary timing seizures to the end of a rotation window reads not the phase mean but the
per-rotation maximum (\S\ref{sec:threat}). Recomputing the pretrained victim's usable fraction as
that maximum raises it negligibly at the fastest rotation and sharply as $R$ grows: at $R{=}5$\,s on
AT\&T the protective corner holds (CIFAR-100 unchanged at $0.000$ over $24$ windows, CIFAR-10 only
$0.118\!\rightarrow\!0.333$), but at $R{=}10$\,s over $12$ windows CIFAR-10 climbs
$0.472\!\rightarrow\!0.917$ and CIFAR-100 $0.008\!\rightarrow\!0.083$. On T-Mobile at $R{=}5$\,s
CIFAR-10 rises $0.486\!\rightarrow\!0.853$ over $95$ windows. The mean understates a timing adversary everywhere
but the fast-rotation corner a defender would operate in.

Two checks bound the envelope. The bar sweep from $0.6$ to $0.95$, run on both victims, shows the
passive ordering above is a $\tau{=}0.8$ statement: sharing the byte axis, each envelope turns on
which exit a seizure must reach, so the two coincide at every swept bar but $\tau{=}0.8$ on both
datasets, and a stricter bar only steps the fraction down without lifting the adversary over it: at
$R{=}30$\,s the from-scratch usable fraction falls from $0.942$ to $0.809$ on CIFAR-10 and $0.805$
to $0.380$ on CIFAR-100 as $\tau$ rises from $0.8$ to $0.9$. A prefix-constrained optimal scheduler (an ILP,
EWMA $\alpha{=}0.3$, $\beta{=}2.0$, deadline $R$) never beats greedy across $180$ matched pairs on
the pretrained victim, trailing by at most $1.15$\,pp of mean $a/a_{\max}$ and $3.54$\,pp of usable
fraction, so greedy is a tight upper bound and the envelope no scheduler artifact. Every number here is the naive passive baseline that \S\ref{sec:results-thief} recomputes
against a capable adversary.

\subsection{The public prior and the collapse of the envelope}
\label{sec:results-thief}
The threat model concedes the architecture to the adversary (\S\ref{sec:threat}). Once the defender
also builds on public weights, the stages a seizure fails to capture stop being an unknown
the adversary has to learn: they are torchvision ImageNet-1k ResNet-18 layers $k{+}1$ through $4$,
and the adversary can download them. We measure what that is worth. The attack seizes units $0..k$,
that is the chunk prefix $c_1\dots c_{k+1}$, freezes it, and fine-tunes the withheld stages on $N$ labelled in-distribution
samples for $S$ gradient steps, exactly the procedure fixed by a pre-registration dated
2026-08-01 whose outcome branches were committed before the results existed; the initialisation
is its only changed component, and an $S{=}0$ gate reproduced the passive exit accuracy to
$|\Delta|{=}0.000000$ before any result was read. The grid is
$2$ datasets $\times$ $4$ depths $\times$ $N\!\in\!\{1000, 5000, 50000\}$ $\times$
$S\!\in\!\{100, 500, 2000\}$ $\times$ $3$ seeds, so each of the $216$ cells is one seed-level run.
A cell counts as cleared only when the adversary lifts a seizure across the usability bar that the
passive prefix left below it, not where the prefix was already usable; $81$ cannot clear by
construction, their passive prefix already above the bar, so of the $135$ eligible cells $50$ clear.

\begin{figure}[tb]
\centering
\includegraphics[width=0.9\columnwidth]{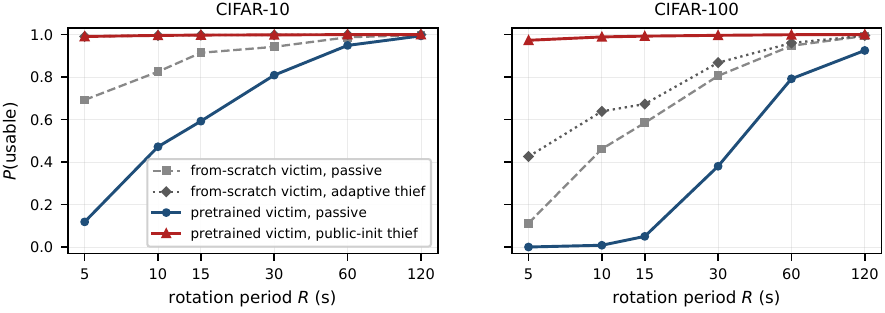}
\caption{The capture envelope collapses once the withheld stages have a public origin.
Usable-seizure fraction $P(\text{usable})$ against rotation period $R$; AT\&T trace,
$\tau{=}0.8\,a_{\max}$ of the regime plotted. Series, in plot order: from-scratch victim, passive;
from-scratch victim, adaptive thief; pretrained victim, passive; pretrained victim, public-init
thief.}
\label{fig:collapse}
\end{figure}

Table~\ref{tab:thief} gives recovered accuracy (top-1 at the deepest exit, held-out split) at
$N{=}50{,}000$, $S{=}2000$. Every seizure
depth lands at or above $90\%$ of $a_{\max}$ on both datasets, including $k{=}0$, where the seizure
caught the first chunk and nothing else. Seizure depth no longer helps the defender, and that is a
change of mechanism rather than of degree. Against the from-scratch victim, an adversary that had to
learn the withheld stages from random initialisation gained $+48.5$\,pp at $k{=}0$ and
$-0.61$\,pp at $k{=}3$ on CIFAR-10, and $+42.4$ down to $-3.17$\,pp over the same depths on
CIFAR-100, at $N{=}50{,}000$ and $S{=}5000$; a deeper seizure handed it a stronger trained exit to
beat. It alone ends the from-scratch regime's protection on CIFAR-10, lifting $P(\text{usable})$
from $0.692$ to $0.991$ at $R{=}5$\,s and $0.942$ to $0.998$ at $R{=}30$\,s, while on
CIFAR-100 rotation still separates $0.426$ from $0.868$: learning the withheld
stages costs an adversary what downloading them does not.
Carried into the pretrained victim's envelope (Fig.~\ref{fig:collapse}, public-init thief series),
the public prior takes the CIFAR-100 protective corner from zero usable seizures to $97.3\%$, and
CIFAR-10 from $11.8\%$ to $99.1\%$, with no rotation on AT\&T leaving the defender better placed.
Against this adversary, rotation no longer bounds anything a defender would trade accuracy and
energy for.

The obvious alternative explanation is that fine-tuning simply produced an easier victim. It did
not. We re-ran the identical attack against the identical pretrained frontier with the withheld
stages randomly initialised, at the same budgets and seeds: $2$ of the $135$ eligible cells clear the
bar against $50$ for the public initialisation. At $N{=}50{,}000$, $S{=}2000$ the public prior is worth $+21.9$
and $+17.1$\,pp over random initialisation at $k{=}0$ and $k{=}1$ on CIFAR-10 and $+33.7$ and
$+24.0$\,pp on CIFAR-100, and nothing at $k{=}3$, where the captured prefix already sits near
$a_{\max}$ and neither route has anything left to recover. At the envelope level on CIFAR-10 the
random-initialised adversary moves $P(\text{usable})$ by $+0.00$\,pp at every rotation on both
traces, while the public-initialised one moves it by up to $+87.3$\,pp. The CIFAR-100 envelope row
turns on the knife-edge cell Table~\ref{tab:thief}'s caption flags, so it is not used; the control's
CIFAR-100 evidence is the cell counts and shallow-depth gaps. The public prior causes the collapse;
the victim does not.

The collapse does not need the same data on both tasks, and that difference is the sharpest limit on
the result. Capping the adversary at $N\!\le\!5000$ labelled samples, a tenth of the training set,
leaves CIFAR-10 untouched: $P(\text{usable})$ at $R{=}5$\,s is still $11.8\%\!\rightarrow\!99.1\%$,
because the adversary reaches $0.8112$ at $c_1$ and $0.8228$ at $c_2$ against a $0.7542$ bar. The
same cap removes the CIFAR-100 collapse entirely, moving the envelope $+0.00$\,pp at every rotation,
because there the adversary reaches only $0.4560$ and $0.4709$ against a $0.6181$ bar. The CIFAR-100
result therefore requires the victim's full $50{,}000$-image labelled training set. Throughout, the
adversary is handed in-distribution labelled data and the correct architecture, which a
field-capture adversary plausibly lacks, so these recoveries are upper bounds on a realistic attack,
and CIFAR-10 is the collapse that survives a realistic data budget.

That public weights help an adversary is not a new observation; its size, shape and cost are what we
measure. The public prior is an initialisation, not an approximation: the victim's fine-tuned stages
sit at mean cosine $0.494$ (CIFAR-10) and $0.361$ (CIFAR-100) from public per output filter, $99.8\%$
below $0.9$, so the adversary relearns equally good weights rather than recovering the victim's.
Delivery-side secrecy and public-backbone deployment are therefore incompatible, and withholding or
obfuscating different stages does not repair it: the adversary still starts from public weights.

\begin{table}[tb]
\centering
\caption{Pretrained victim, $N{=}50{,}000$, $S{=}2000$, by seizure depth $k$: passive is the
accuracy the seized prefix serves at its own exit, the random- and public-init columns recovered
accuracy at the deepest exit after fine-tuning the withheld stages;
$a_{\max}{=}0.9428$ and $0.7726$, bars $0.7542$ and $0.6181$.
The random-init control at CIFAR-100 $k{=}2$ clears its bar by
$0.0003$ ($1$ of $3$ seeds)---a knife-edge, not an effect---so the control's evidence is the cell
counts and shallow-depth gaps in the text.}
\label{tab:thief}
\setlength{\tabcolsep}{4pt}
\footnotesize
\begin{tabular}{llcccc}
\toprule
frontier & $k$ & passive & random init & public init & of $a_{\max}$ (\%) \\
\midrule
\multirow{4}{*}{CIFAR-10}  & 0 & 0.3286 & 0.6873 & \textbf{0.9065} & 96.2 \\
                           & 1 & 0.7414 & 0.7247 & \textbf{0.8956} & 95.0 \\
                           & 2 & 0.9028 & 0.8941 & 0.9048 & 96.0 \\
                           & 3 & 0.9397 & 0.9381 & 0.9368 & 99.4 \\
\midrule
\multirow{4}{*}{CIFAR-100} & 0 & 0.0960 & 0.3736 & \textbf{0.7101} & 91.9 \\
                           & 1 & 0.3216 & 0.4645 & \textbf{0.7045} & 91.2 \\
                           & 2 & 0.6033 & 0.6184 & \textbf{0.6963} & 90.1 \\
                           & 3 & 0.7474 & 0.7304 & 0.7355 & 95.2 \\
\bottomrule
\end{tabular}
\end{table}

\subsection{Validation against measured TCP delivery}
\label{sec:results-mmlink}
Section~\ref{sec:results-capture} obtains delivered bytes by integrating the raw trace, which assumes
the link hands over bytes at trace capacity; a real transport delivers less. We measured the
shortfall through mm-link, which releases packets into a real network namespace at the trace's
instantaneous rate. A bulk CUBIC flow runs into it, and the receiver's timestamped
cumulative-delivery curve replaces the trace integral in the same accounting, with frontier,
rotation grid, usability bar and seizure enumeration held fixed, so any difference is transport
alone (AT\&T, $3$ repeats per cell). Every cell falls below the integral in mean $a/a_{\max}$. A warm
flow held open for the whole trace, with rotations applied offline (an upper bound on any transport,
since it never repays slow start), sits $-0.78$ to $-5.03$\,pp below it across individual repeats, and a fresh
connection per rotation epoch sits further below, $-2.49$ to $-11.60$\,pp over repeats, its
deepest per-dataset cell means $-11.02$\,pp on CIFAR-10 at $R{=}5$\,s and $-8.38$\,pp on CIFAR-100 at $R{=}10$\,s, because it repays slow
start once per epoch. Slow start bounds the extremes without ordering the cells: the shortfall is
not monotone in $R$, and the CIFAR-100 warm flow deepens from $-2.00$ to $-3.32$\,pp as $R$ rises
from $5$ to $30$\,s. The published envelope is therefore
conservative in the defender's favour under either reading, though this validation covers one trace,
one $200$-packet droptail queue, CUBIC, and the CIFAR frontiers only.

\subsection{Residency and exchange}
\label{sec:results-resident}
The bound has a price, measured on the pretrained victim. Pinning a chunk permanently resident to
serve through a link outage is a step function whose step is the usability bar.
On AT\&T at $R{=}30$\,s with a $30$\,s outage, pinning nothing, $c_1$, or $c_2$ leaves CIFAR-10
availability at $0.3579$ and CIFAR-100 at $0.1135$; availability reaches $1.0000$ only at $c_3$
(CIFAR-10) and $c_4$ (CIFAR-100), where the accuracy floor conceded permanently to every seizure is
$0.9576$ and $0.9674$ of $a_{\max}$. Since the bar rose with
$a_{\max}$, the prefix that carries a device through an outage hands an adversary a model as good as
the device's own.

Tightening the envelope charges the device instead. Under the analytic LTE energy model (receive
power $1.5$\,W, plus $10$\,J per rotation for connection setup, the fresh-connection term of
\S\ref{sec:results-mmlink}), moving CIFAR-100 on AT\&T from $R{=}30$\,s to $R{=}5$\,s takes passive
usable recovery from $0.380$ to $0.000$, at a cost of delivery energy $6600$ to $12600$\,J per hour
($1.8$ to $3.5$\,W continuous), entirely the per-rotation setup term, and device mean accuracy
$0.761$ to $0.354$ of $a_{\max}$, while against an adversary holding the public prior the same spend
barely moves usable recovery ($0.996$ to $0.973$).

\section{Limitations and Open Questions}
\label{sec:discussion}
Several limitations bound the claims. First, and the primary open question: the victim's fine-tuned
stages end far from their public origin (\S\ref{sec:results-thief}), yet that prior still carries an
adversary to $\ge\!90\%$ of $a_{\max}$. Why does so distant a starting point still work, and what
adaptation would defeat it? We measure neither, and no delivery-side control reaches either; this is
a question about training, not rotation. Second, the CIFAR-100 collapse needs the victim's
full $50{,}000$-image labelled training set and does not occur at $5000$ samples, whereas the
CIFAR-10 collapse still occurs in full at that budget; the adversary is handed in-distribution
labels and the correct architecture throughout, which a field-capture adversary plausibly lacks.
Third, the from-scratch victim's depth-decay gradient is measured on an asymmetric grid: $k{\ge}1$
was run only to $S{=}5000$ steps and $k{=}0$ to $S{=}20{,}000$, where the $k{=}0$ advantage still
climbed to $+51.7$\,pp on CIFAR-10. Fourth, the evaluation is CIFAR-scale image classification only,
and covers no language or generative model. Fifth, we evaluate no physical edge hardware: traces are
replayed and inference is measured on a server GPU. Sixth, the energy model is analytic, with the
stated constants, rather than a measured power trace. Finally, the $P(\text{full}){=}0$ result
holds on AT\&T alone, up to $R{=}60$\,s, and is weakened by the flat $c_4\!\to\!c_5$ segment of
the from-scratch frontier.

\section{Conclusion}
\label{sec:conclusion}
When an image classifier is streamed into a field node's volatile memory and never persisted, a
physical seizure recovers only the prefix resident at that instant; the cost of recovering the
rest depends on what that remainder is. Against a from-scratch victim the remainder must be
learned: the adaptive adversary gains most at shallow seizures, ends the CIFAR-10 envelope at
every rotation we test, and stays rotation-bounded on CIFAR-100. Against
the same architecture sliced from a public pretrained backbone the remainder is downloaded and
rotation bounds nothing, a collapse needing the victim's entire labelled training set on
CIFAR-100 and a tenth of it on CIFAR-10. Delivery-side secrecy raises the price of a seizure; a
public copy of the withheld part reduces it to nothing. Yet the relearned stages end far from their
public origin, and the prior works. Why, and what adaptation would defeat it? We measure
neither; no delivery-side control reaches either.

\let\OLDthebibliography\thebibliography
\renewcommand\thebibliography[1]{\OLDthebibliography{#1}%
  \setlength{\parskip}{0pt}\setlength{\itemsep}{0pt plus 0.3ex}}
\bibliographystyle{IEEEtran}
\bibliography{ref}

\end{document}